\documentclass[12pt]{article}
\usepackage{amssymb}
\usepackage{amsbsy}
\usepackage{epsfig}
\usepackage{verbatim}
\makeatletter \@addtoreset{figure}{section}
\def\thefigure{\thesection.\@arabic\c@figure}
\def\fps@figure{h, t}
\@addtoreset{table}{bsection}
\def\thetable{\thesection.\@arabic\c@table}
\def\fps@table{h, t}
\@addtoreset{equation}{section}

\newtheorem{corollary}{Corollary}[section]
\newtheorem{definition}{Definition}[section]
\newtheorem{theorem}{Theorem}[section]
\newtheorem{proposition}{Proposition}[section] 

\newtheorem{examps}{Examples}[section]

\newtheorem{lemma}{Lemma}[section]
\newtheorem{remark}{Remark}[section]
\newtheorem{remarks}[remark]{Remarks}

\def\bx{\begin{example}}
\def\ex{\end{example}}
\def\bxs{\begin{examps}. \rm\begin{enumerate}}
\def\exs{\end{enumerate}\end{examps}}
\def\bd{\begin{definition}}
\def\ed{\end{definition}}
\def\bt{\begin{theorem}}
\def\et{\end{theorem}}
\def\bp{\begin{proposition}\rm}
\def\ep{\end{proposition}}
\def\bc{\begin{corollary}}
\def\ec{\end{corollary}}
\def\bl{\begin{lemma}\em}
\def\el{\end{lemma}}
\def\be{\begin{equation}}
\def\ee{\end{equation}}
\def\br{\begin{remark}\rm\small}
\def\er{\end{remark}}
\def\brs{\begin{remarks}.\\ \rm\
\begin{enumerate}}
\def\ers{\end{enumerate}\end{remarks}}
\def\bea{\begin{eqnarray}}
\def\eea{\end{eqnarray}}

\def\ra{{\rightarrow}}

\def\wt{\widetilde}

\def\tr{\mathrm {tr}}
\def\det{\mathrm {det}}

\def\&{&{\hskip -20pt}}

\def\HH{\mathcal{H}}
\def\II{\mathcal{I}}
\def\KK{\mathcal{K}}
\def\NN{\mathcal{N}}

\def\WW{\mathcal{W}}

\def\Cb{\mathbf{C}}

\def\Nb{\mathbf{N}}

\def\Rb{\mathbf{R}}

\def\Zb{\mathbf{Z}}

\def\Hbb{\mathbb{H}}

 \def\grg{\mathfrak{g}}

 \def\gro{\mathfrak{o}}

 \def\gru{\mathfrak{u}}

\def\grGl{\mathfrak{Gl}} 
 \def\grsp{\mathfrak{sp}}

\def\nchi{\hbox{\raise 2.5pt\hbox{$\chi$}}}

\date{}

\begin{document}
\baselineskip 16pt 
\begin{flushright}
CRM-3250 (2007)
\end{flushright}
\medskip
\begin{center}
\begin{Large}\fontfamily{cmss}
\fontsize{17pt}{27pt}
\selectfont
\textbf{Determinantal identity for multilevel systems and finite determinantal point processes}
\footnote{Work of J.H. supported in part by the Natural Sciences and Engineering Research Council of Canada (NSERC) and the Fonds FCAR du Qu\'ebec.  Work of A.O. supported by joint RFBR Consortium EINSTEIN grants Nos. 06-01-92054; 05-01-00498 and RAS Program ``Fundamental Methods in Nonlinear Physics''.}
\end{Large}\\
\bigskip
\begin{large}  {J. Harnad}$^{\dagger \ddagger}$\footnote{harnad@crm.umontreal.ca}
 and {A. Yu. Orlov}$^{\star}$\footnote{ orlovs@wave.sio.rssi.ru}
\end{large}
\\
\bigskip
\begin{small}
$^{\dagger}$ {\em Centre de recherches math\'ematiques,
Universit\'e de Montr\'eal\\ C.~P.~6128, succ. centre ville, Montr\'eal,
Qu\'ebec, Canada H3C 3J7} \\
\smallskip
$^{\ddagger}$ {\em Department of Mathematics and
Statistics, Concordia University\\ 7141 Sherbrooke W., Montr\'eal, Qu\'ebec,
Canada H4B 1R6} \\ 
\smallskip
$^{\star}$ {\em Nonlinear Wave Processes Laboratory, \\
Oceanology Institute, 36 Nakhimovskii Prospect\\
Moscow 117851, Russia } \\
\end{small}
\end{center}
\bigskip
\bigskip
\begin{center}{\bf Abstract}
\end{center}
\smallskip

\begin{small}
We give a simple algebraic derivation of a useful determinantal identity
for multilevel systems such as random matrix chains and finite determinantal
point processes, with applications to the  calculation of point correlators, gap probabililties 
and Janossy densities.
\bigskip
\end{small}
\bigskip \bigskip

\section{Multilevel determinantal ensembles}
\subsection {Joint probability distributions }
 
   The following common setting underlies  various multilevel determinantal
ensembles, including chains of random matrices \cite{EM, H, S}, and finite
determinantal point processes such as  Dyson processes \cite{TW2} and polynuclear growth \cite {PS}.
Let $\{(\Gamma_j, d\mu_j)\}_{j=1 \dots m}$ be a set of measure spaces and
$\{\tilde{H_j}:= L^2(\Gamma_j, d\mu_j)\}_{j=1 \dots m}$ the Hilbert spaces
of square integrable functions on them.
 Suppose we are given $m -1$ functions  $\{w_{j+1, j}\}_{j=1\dots m-1}$  on
the product spaces $\Gamma_{j+1}\times \Gamma_{j}$, such that the 
corresponding integral operators
\bea
w_{j+1, j}: \hat{H}_j &\&\ra  \hat{H}_{j+1} \cr
w_{j+1, j}(f)(x^{(j+1)}) :=&\&  \int_{\Gamma_j} w_{j+1,j}(x^{(j+1)},
x^{(j)})f(x^{(j)})d\mu_j(x^{(j)}),
\eea
together with their transposes
\bea
w^*_{j+1, j}: \hat{H}_{j+1} &\&\ra  \hat{H}_j \cr
w^*_{j+1, j}(f)(x^{(j)}) :=&\&  \int_{\Gamma_{j+1}} w_{j+1,j}(x^{(j+1)},
x^{(j)})f(x^{(j+1)})d\mu_j(x^{(j+1)}),
\eea
are well defined injective maps on a sequence of dense subspaces
$\hat{H}_1\subset \wt{H}_1$ , $\hat{H}_2 = w_{21}(\hat{H}_1)\subset\wt{H}_2$,
$\dots,$
$\hat{H}_m =  w_{m,m-1}(\hat{H}_{m-1})\subset\wt{H}_m$, as are their composites:
\be
w_{k j} := w_{k,k-1} \circ \dots \circ w_{j+1,j}, \quad  m\ge k > j \ge 1.
\label{wij}
\ee
Let $H_1\subset \hat{H}_1$ be an $N$ dimensional subspace with basis 
$\{\psi_a^{(1)}\}_{a=1, \dots N}$, and  $\{H_j:= w_{j1}(H_1)\subset \wt{H}_j\}_{j=1, \dots , m}$,  the corresponding subspaces obtained by applying the operators $\{w_{j1}\}$ to $H_1$, with bases 
\be
\{\psi_a^{(j)}:= w_{j1}(\psi_a^{(1)})\}_{a=1, \dots N}.
\label{psiaj}
\ee
Now choose the basis $\{\phi_a^{(m)}\}_{a=1, \dots N}$  for  $H_m$ dual to 
$\{\psi_a^{(m)}\}_{a=1, \dots N}$, and let 
\be
\phi_a^{(j)}:= w^*_{mj}(\phi_a^{(m)}), \quad a=1, \dots N
\label{phiaj}
\ee
be the corresponding dual bases for the $H_j$'s. Thus 
\be
\int_{\Gamma_j} \psi_a^{(j)}(x^{(j)}) \phi_b^{(j)}(x^{(j)}) d\mu_j(x^{(j)})= \delta_{ab}.
\label{dualphipsi}
\ee

Assume that
\be
\det(\psi^{(1)}_a(x^{(1)}_b))
\det(\phi^{(m)}_a(x^{(m)}_b))\prod_{j=1}^{m-1} \det (w_{j+1, j}(x^{(j+1)}_a,
x^{(j)}_b)) \prod_{j=1}^m\prod_{a=1}^N d\mu_j (x_a^{(j)})
\ee
is a  positive Borel measure on $\prod_{j=1}^m (\Gamma_j)^ N$.
It follows from the duality relations (\ref{dualphipsi}) that this is a  normalized
probability measure
\bea
\prod_{j=1}^m\left(\int_{\Gamma_j^N}\prod_{a=1}^N \left(d\mu_j (x_a^{(j)})\right)
\right)&\& \det(\psi^{(1)}_a(x^{(1)}_b))\det(\phi^{(m)}_a(x^{(m)}_b)) \cr
&\&  \times\prod_{j=1}^{m-1} \det (w_{j+1, j}(x^{(j+1)}_a, x^{(j)}_b)) 
 =1
\eea
so that
\be
P^{N,m} (x^{(j)}_a) :=\det(\psi^{(1)}_a(x^{(1)}_b)) \det(\phi^{(m)}_a(x^{(m)}_b))\prod_{j=1}^{m-1} \det (w_{j+1, j}(x^{(j+1)}_a, x^{(j)}_b))  
\label{PNm}
\ee
may be interpreted as a joint probability density for points in $\prod_{j=1}^m (\Gamma_j)^ N$.

A typical example of such probability measures arises in the theory of random matrices.
We start with a chain of $N \times N$ Hermitian matrices $\{M_j \in \Hbb^{N}\}_{j=1, \dots m}$
with  nearest neighbor couplings of exponential type $e^{\tr(M_j M_{j+1})}$, and each
site in the chain contributing a conjugation invariant factor $e^{-\tr (V_j(M_j))}$ to the joint probability density (where the $V_j$'s are e.g. polynomials in the matrices), and  the product gives a normalizable
positive measure.  Integrating over the ``angular variables'',  using the Harish-Chandra-Itzykson-Zuber 
(HCIZ) identity  \cite{IZ}, leads to a reduced joint probability measure of the form ( \ref{PNm}) on the space of eigenvalues,  where the $\Gamma_j$'s are all taken as the real line,
\bea
w_{j+1,j}(x^{(j+1)}, x^{(j)}) &\& :=e^{x^{(j)}x^{(j+1)} -{1\over 2}(V_{j}(x^{(j)}) + V_{j+1}(x^{(j+1)}))},\\
\psi^{(1)}_a(x^{(1)}):= p_{a-1}(x^{(1)}) e^{-{1\over 2} V_1(x^{(1)})} ,  &\& \quad
\phi^{(m)}_a(x^{(m)}) := s_{a-1}(x^{(m)}) e^{-{1\over 2} V_m(x^{(m)})}
\label{psi1phimpolynom}
\eea
and $\{p_a(x^{(1)}), s_a(x^{(m)})\}_{a=0, 1, \dots}$ are a sequence of pairs of polynomials of degrees $a$ satisfying the biorthogonality relations
\be
 \prod_{j=1}^m\left(\int_{\Gamma_j^N} d\mu_j (x^{(j)})\right)
\psi_a^{(1)}(x^{(1)})\phi_b^{(m)}(x^{(m)}) \prod_{j=1}^{m-1}w_{j+1, j}(x^{(j+1)}, x^{(j)}))
=\delta_{ab} .
\label{biorthog}
\ee

If all the $\Gamma_j$'s are identified as the same space $\Gamma$,  (\ref{PNm}) may also be interpreted as a measure on the space of $m$-step paths  of $N$-tuples of points in $\Gamma$,
defining a point process, in which $j$ is viewed as a discrete time parameter. Examples of this kind lead to Dyson processes \cite{TW2}, describing diffusion of eigenvalues, and Polynuclear Growth \cite{PS}.

\subsection {Point correlators and gap probabilities}

  What characterizes such multilevel determinantal ensembles is that multi-point correlators
(marginal distributions), gap probabilities and expectation values may all be expressed
as  determinants, either finite,  or determinants of Fredholm integral 
operators, in terms of a single $m\times m$ matrix kernel function 
$\check{K}_{ij} (x^{(i)}, x^{(j)})_{i,j, = 1, \dots m}$ defined as follows:
\be
\check{K}_{ij}(x^{(i)} x^{(j)}) := K_{ij}(x^{(i)}, x^{(j)}) - w_{ij}(x^{(i)},x^{(j)}),
\label{Kcij}
\ee
where
\bea
K_{ij}(x^{(i)}, x^{(j)}) := \sum_{a=1}^N \psi_a^{(i)}(x^{(i)}) \phi_a^{(j)}(x^{(j)}) .
\label{Kij} \\
w_{ij}(x^{(i)}, x^{(j)}) := 0  \quad {\rm if \quad} i\le j
\eea
and $w_{ij}(x^{(i)}, x^{(j)})$ is defined as in (\ref{wij}) for $i > j$.
In terms of this, the joint probability density (\ref{PNm}) may be expressed as 
\be
P^{N,m} (x^{(j)}_a) = \det (\check{K}_{ij}(x^{(i)}_a, x^{(j)}_b))
\label{PNm_detKK}
\ee
where $\check{K}_{ij}(x^{(i)}_a, x^{(j)}_b)$ is viewed as the $((i,a),(j,b))$
element of a matrix of dimension $Nm \times Nm$, labelled by pairs of double
indices $1 \le i,j \le m, \ 1\le a,b \le N$.
By  integrating eq. (\ref{PNm_detKK}) over some of the  variables while setting the rest equal to 
fixed values, it follows \cite{EM} that the correlation function giving the probability density 
for finding $k_j$ elements in $\Gamma_j$ at the points $\{x^{(j)}_1, \dots x^{(j)}_{k_j}\}$ for
$j=1 \dots m$ is similarly given, within a combinatorial factor,  by the  $\sum_{j=1}^m k_j \times \sum_{j=1}^m k_j$  determinant
\be
P^{N,m}_{k_1,\dots k_m} (\{x^{(j)}_1, \dots x^{(j)}_{k_j}\}_{j=1 \dots m}) :=\det
(\check{K}_{ij}(x^{(i)}_a, x^{(j)}_b))\vert_{{1\le a \le k_i  \atop 1\le b
\le k_j}}\ .
\label{Pk1km}
\ee
Alternatively, choosing a measurable subset $J_j \subset \Gamma_j$ of each $\Gamma_j$,
the probabability E(0, {\bf J}) of finding no points within the set 
\be
{\bf J} := J_1 \times J_2 \times \dots J_m
\ee
is given, within a normalization constant, by the  Fredholm determinant  (\cite{S}, \cite{TW2})
\be
E^{N,m}(0, {\bf J}) =C_{Nm} \det (\II - \check{\KK} \circ \chi_{\bf J}),
\label{ENmJ}
\ee
where $\check{\KK}$ is the $m\times m$ matrix integral operator with kernel
$\check{K}_{ij}$ defined in (\ref{Kcij}) and $\chi_{\bf J}$ is the direct sum of
the operators of multiplication by the characteristic function $\{ \chi_{J_j}\}_{j=1, \dots m}$
on each of the factors in the direct sum
\be
\hat{H}:= \oplus_{j=1}^m \hat{H}_j. 
\label{directsum}
\ee
More refined statistics, giving the probability of finding any specified number of points within 
disjoint subintervals $\{J_{jl}\}_{l=1, m_j}$ of the $J_j$'s may be similarly computed by replacing the characteristic functions $\chi_{J_j}$ by weighted ones $\sum_{l=1}^{m_j} z_{jl} \chi_{J_{jl}}$ 
and evaluating the coefficients of the monomials $\prod_{j=1}^m \prod_{l=1}^{m_j} z_{jl}^{k_{jl}}$.

A determinantal expression similar to (\ref{Pk1km}) may also be found for the so-called ``Janossy densities''  \cite{H, S} , which give the probabilities of finding elements at the given set of points  
$\{x^{(j)}_1, \dots x^{(j)}_{k_j}\}$ within the subsets  $(J_1, J_2,  \dots J_m)$, while all others are 
on the outside, normalized to the probability of all points being on the outside. 
In that case, the  integral kernel $\check{K}_{ij}(x^{(i)}_a, x^{(j)}_b)$ appearing in (\ref{Pk1km}) 
needs simply to be 
replaced by  the corresponding kernel $\check{R}^{\chi_{\bf J}} _{ij}(x^{(i)}_a, x^{(j)}_b)$ 
of the Fredholm resolvent operator:
\be
R^{\chi_{\bf J}} := ({\bf 1}- \check{\KK}^{\chi_{\bf J}})^{-1} \circ \check{\KK}^{\chi_{\bf J}} , 
\ee
where
\be
\check{\KK}^{\chi_{\bf J}}:=\check{\KK}\circ \chi_{\bf J}.
\ee

  These results may  all be obtained as consequences of a single  determinantal  identity for  multilevel ensembles, which   can be expressed in a purely algebraic form. 
  In the next section this algebraic identity will be stated, and a very simple proof given. 
All the aboveexpressions for  correlators and gap probabilities  may  be deduced
from it, as can similar expressions for arbitrary multilevel determinantal ensembles.

    This  result is not at all new; it was derived e.g. in refs. \cite{EM}, \cite{H}, \cite{S}, \cite{TW2}
 in a number of particular cases, using a variety of different methods.
       The point of giving a new proof of the underlying identity is just to unify and simplify the argument,
showing its universal applicability. By suitably particularizing to the various cases of interest, 
all previous results of this type follow, and determinantal expressions may be similarly 
derived for further cases that have not previously been studied.  Some examples, consisting
of chains of Lie algebra valued matrices, including those having various nearest neighbour couplings that admit reductions similar to that of Harish-Chandra-Itzykson-Zuber, are  given in section 3.
  
  Following Tracy and Widom  \cite{TW1, TW2}, the relevant integrals are first obtained in 
  a uniform way by introducing a set of $m$ measurable functions or distributions
  $\{\rho_1, \dots ,\rho_m\}$ on $\{\Gamma_1, \dots ,\Gamma_m\}$ and considering the
  integral obained by replacing each $d\mu_j$ by $d\mu_j (1-\rho_j)$.
  \bea
    P^{N,m}_{\rho_1, \dots, \rho_m}:= &\&
\prod_{j=1}^m\left(\int_{\Gamma_j^N} \prod_{a=1}^N\left(d\mu_j (x_a^{(j)})  (1-\rho_j(x_a^{(j)})
\right)\right)
\det(\psi^{(1)}_a(x^{(1)}_b))\det(\phi^{(m)}_a(x^{(m)}_b)) \cr
&\& \qquad \quad   \times\prod_{j=1}^{m-1} \det (w_{j+1, j}(x^{(j+1)}_a, x^{(j)}_b)) 
\label{Prho1rhom}
  \eea
  We then apply a multi-level version of the Andreief identity \cite{A} to express this integral as
  an $N\times N$ determinant.
  \be
     P_{\rho_1, \dots, \rho_m}: = (N!)^{m} \det G,
     \ee
     where
     \be
     G_{ab} := 
\prod_{j=1}^m\left(\int_{\Gamma_j}\left(d\mu_j (x^{(j)})  (1-\rho_j(x^{(j)})
\right)\right)
\psi^{(1)}_a(x^{(1)}) \phi^{(m)}_b(x^{(m)}) \prod_{j=1}^{m-1} w_{j+1, j}(x^{(j+1)}, x^{(j)}) .
\ee
Like the one-level version, this  follows easily from the invariance of the integrand under permutations in the various integration variables at each level.

  By choosing the factors $\{\rho_j\}$ in different ways, Tracy and Widom \cite{TW1, TW2} showed that  point correlation functions, gap probabilities and Janossy densities may all be obtained   from the formula  (\ref{Prho1rhom}). For example,  choosing the $\rho_j$'s as:
  \be
  \rho_j:=-\sum_{l=1}^{k_j}z_{jl} \delta_{X_l^{(j)}}
  \ee
the coefficient of the monomial factor $\prod_{j=1}^m\prod_{l=1}^{k_j}z_{jl}$
  in $ P_{\rho_1, \dots, \rho_m}$ gives the point correlator  $P_{k_1,\dots k_m} (\{X^{(j)}_1, \dots X^{(j)}_{k_j}\}_{j=1 \dots m}) $. Choosing instead
   \be
  \rho_j:= \sum_{l=1}^{k_j}z_{jl} \chi_{{}_{X_{J_{jl}}}}
  \ee
  gives the generating function for probabilities of finding the various numbers of
  points within disjoint subintervals $J_{jl} \subset J_j$.  Choosing 
   \be
  \rho_j:= \chi_{{}_{J_j}}-\sum_{l=1}^{k_j}z_{jl} \delta_{X_j^l}
  \ee
  and evaluating the coefficient of the monomial factor $\prod_{j=1}^m\prod_{l=1}^{k_j}z_{jl}$
  in $ P_{\rho_1, \dots, \rho_m}$ gives the corresponding Janossy densities.
 
In order to apply the abstract form of the determinantal identity derived in the following section, it is convenient to interpret the matrix elements of $G$ as  scalar products. To express integrals involving the functions $\psi_a^{(j)}$ and
$\phi_a^{(j)}$ as scalar products, these will be represented as vectors or linear forms,
using the bra and ket notation:  $<\psi_a^{(j)}|$, $<\phi_a^{(j)}|$, $|\psi_a^{(j)}>$ and $|\phi_a^{(j)}>$.
Thus, $G_{ab}$ may equivalently be written as:
\be
G_{ab} = <\psi^{(1)}_a| (1-\rho_1) w^*_{2,1} (1-\rho_2 )\cdots  (1-\rho_{m-1})w^*_{m+1,m}
(1-\rho_m) |\phi^{(m)}_b>
\label{Gab}
\ee
where 
\be
\rho_j:\hat{H}_j \ra \hat{H}_j
\ee
is the operator of multiplication by $\rho_j$, which is symmetric ($\rho_j =\rho_j^*$)
with respect to the scalar product
\be
<v^{(j)} | u^{(j)}>_j:=  \int_{\Gamma_j} v^{(j)}(x^{(j)})u^{(j)}(x^{(j)} )d\mu_j(x^{(j)})
\ee
on $\hat{H}_j$.
 
\section{Multilevel determinantal identity}

To provide a uniform algebraic setting for all the previous examples, we assume 
that we have $m$ Euclidean inner product  spaces $\{\hat{H}_1, \dots, \hat{H_m}\}$, and 
denote their direct sum by $\hat{H}$ as in  (\ref{directsum}). We also assume  that there 
are $m-1$ injective linear maps:
\be
w_{j+1, j}: \hat{H}_j \ra  \hat{H}_{j+1}, \quad j=1, \dots, m-1
\ee
with their duals
\be
w^*_{j+1, j}: \hat{H}_{j+1} \ra  \hat{H}_j 
\ee
as well as $m$  endomorphisms of the spaces $\hat{H}_j$
\be
\rho_j: \hat{H}_j \ra  \hat{H}_j
\ee
that are symmetric $\rho_j = \rho_j^*$.  The map $\rho:\hat{H} \ra \hat{H}$ is defined
to be the diagonal one leaving each subspace $\hat{H}_j$ invariant and equal to $\rho_j$ when restricted to $\hat{H}_j$. We also define  composite maps
\be
w_{k j} : \hat{H_j}\ra \hat{H_k}, \quad m\ge k >  j \ge 1
\ee
as in (\ref{wij}), with  $ w_{k j}:=0$ for $k\le j$.  Together, these define an operator 
\be
\WW: \hat{H} \ra \hat{H}
\ee
that is lower triangular with respect to the decomposition (\ref{directsum})
and  can be written in matrix form as
\be
\WW =
\pmatrix{
0 & 0 & 0 & \cdots & 0 &0 \cr
w_{21} & 0 & 0 & \cdots  &  0& 0\cr 
w_{31} & w_{32}  & 0 &  \cdots &0& 0 \cr
\vdots & \vdots & \vdots & \ddots  &\vdots &\vdots \cr
\vdots & \vdots & \vdots & \cdots  & 0 & 0 \cr
w_{m1} & w_{m2} & w_{m3} & \cdots & w_{m, m-1} & 0
}
\ee
This may be expressed as
\be
\WW = \NN \circ (\II-\NN)^{-1}
\ee
where
\be
\NN =
\pmatrix{
0 & 0 & 0 & \cdots & 0&0 \cr
w_{21} & 0 & 0 & \cdots  & 0& 0\cr 
0 & w_{32}  & 0 &  \cdots & 0& 0 \cr
\vdots & \vdots & \vdots & \ddots  &\vdots &\vdots\cr
\vdots & \vdots & \vdots & \cdots  & 0 &0\cr
0 & 0 & 0 & \cdots & w_{m, m-1} & 0
}
\ee
We also denote the corresponding dual upper triangular operators as
\be
\WW^* = 
\pmatrix{
0 & w^*_{21}& w^*_{31} & \cdots & \cdots &w^*_{m1}\cr
0 & 0 & w^*_{32} & \cdots  &  \cdots&w^*_{m2}\cr 
0& 0  & 0 &  \cdots & \cdots& w^*_{m3} \cr
\vdots & \vdots & \vdots & \ddots  &\vdots &\vdots \cr
0&0 & 0 & \cdots  & 0 & w^*_{m, m-1}\cr
0& 0 & 0 & \cdots & 0 & 0
}
\ee
and
\be
\NN^* = 
\pmatrix{
0 & w^*_{21}&0& \cdots & \cdots &0 \cr
0 & 0 & w^*_{32} & \cdots  &  \cdots&0 \cr 
0& 0  & 0 &  \cdots & \cdots& 0  \cr
\vdots & \vdots & \vdots & \ddots  &\vdots &\vdots \cr
0&0 & 0 & \cdots  & 0 & w^*_{m, m-1}\cr
0& 0 & 0 & \cdots & 0 & 0
}
\ee

As previously, let  $H_1\subset \hat{H}_1$ be an $N$ dimensional subspace with basis 
$\{\psi_a^{(1)}\}_{a=1, \dots N}$, and  $\{H_j:= w_{j1}(H_1)\subset \hat{H}_j\}_{j=1, \dots , m}$,  the corresponding subspaces obtained by applying the operators $\{w_{j1}\}$ to $H_1$, with bases 
$\{\psi_a^{(j)}:= w_{j1}(\psi_a^{(1)})\}_{a=1, \dots N})$.
Let $\{\phi_a^{(m)}\}_{a=1, \dots N}$ be the basis  for  $H_m$  dual to  
$\{\psi_a^{(m)}\}_{a=1, \dots N}$, and $\{\phi_a^{(j)}:= w^*_{mj}(\phi_a^{(m)})\}_{a=1, \dots N})\}$
the corresponding dual bases for the $H_j$'s, so that
\be
<\psi_a^{(j)}| \phi_b^{(j)}>= \delta_{ab}, \quad j=1, \dots , m.
\label{dualphipsij}
\ee

 For $a=1, \dots N$,   we introduce  the $m$-component row and column vectors
  \bea
    \Psi_a &\&:= \pmatrix{|\psi_a^{(1)}> \cr \vdots  \cr |\psi_a^{(m)}>}, \qquad   \qquad  \quad
        \Phi_a := \pmatrix{|\phi_a^{(1)}> \cr \vdots  \cr |\phi_a^{(m)}>}  \\\
          \Psi_a^T&\& := (<\psi_a^{(1)}|, \dots ,   <\psi_a^{(m)}|), \quad 
  \Phi_a^T := (<\phi_a^{(1)}|, \dots ,  <\phi_a^{(m)}|)  ,
\eea
which may be viewed as elements of $ \hat{H}$ and $\hat{H}^*$:
\bea
  \Psi_a &\&= \sum_{j=1}^m \iota_j |\psi_a^{(j)}>,  \quad   \Phi_a = \sum_{j=1}^m \iota_j |\phi_a^{(j)}> \cr
   \Psi^T_a &\&= \sum_{j=1}^m  <\psi_a^{(j)}|  \pi_j,
     \quad   \Phi^T_a = \sum_{j=1}^m <\phi_a^{(j)}| \pi_j ,
     \eea
where
\be
\iota_j: \hat{\HH_j}: \ra \oplus_{j=1}^m \hat{\HH}_j    \qquad \pi_j:\oplus_{j=1}^m \hat{\HH}_j \ra \hat{\HH}_j
\ee  
are the standard injection and projection maps.
These may also be expressed as
  \bea
    \Psi_a  &\&=  (\II - \NN)^{-1}\iota_1 |\psi_a^{(1)}> =   (\II + \WW) \iota_1|\psi_a^{(1)}>
       \label{Psia}   
\\
           \Phi_a  &\&=  (\II - \NN^*)^{-1}\iota_m |\phi_a^{(m)}> =   (\II + \WW^*) \iota_m |\phi_a^{(m)}>
               \label{PsiaT}   
\\
       \Psi^T_a  &\&= <\psi_a^{(1)}| \pi_1 (\II - \NN^*)^{-1} =    <\psi_a^{(1)}|\pi_1 (\II + \WW^*)
              \label{Phia}  
         \\
   \Phi^T_a  &\&= <\phi_a^{(m)}| \pi_m (\II - \NN^*)^{-1} =    <\phi_a^{(m)}|\pi_m (\II + \WW^*) .
         \label{PhiaT}  
   \eea
    The duality relations (\ref{dualphipsi}) may equivalently be expressed as
    \be
    <\psi_a^{(1)}|\pi_1 (\II-\NN^*)^{-1}|\phi_b^{(m)}> = \delta_{ab}.
    \label{psiaWphib}
    \ee
  
  In terms of the elements $\Psi_a, \Phi_a \in \hat{H}$, define two maps
\bea
\Psi: &\& \Cb^N \ra \hat{H},  \quad  \Phi: \Cb^N \ra \hat{H}  \cr
\Psi: &\& e_a \mapsto \Psi_a,  \quad \Phi: e_a \mapsto \Phi_a 
\eea
and their transposes 
\bea
\Psi^T : \hat{H} &\&\ra  \Cb^N ,\cr
\cr
\Psi^T:  \pmatrix{ |v^{(1)} >\cr \vdots \cr  |v^{(m)}>}&\& \mapsto \sum_{a=1}^N\sum_{j=1}^m <\psi_a^{(j)}| v^{(j)}> e_a  \\
\cr
\Phi^T : \hat{H} &\& \ra  \Cb^N,   \cr
\cr
\Phi^T:  \pmatrix{ |v^{(1)} >\cr \vdots \cr  |v^{(m)}>} &\& \mapsto \sum_{a=1}^N\sum_{j=1}^m <\phi_a^{(j)}| v^{(j)}> e_a,
\eea
where $\{e_a\}_{a=1,\dots , N}$ is the standard basis for for $\Cb^N$.  

Denote by
\be
\KK:= \Psi \circ \Phi^T 
\ee
 the linear operator $\KK : \hat{H} \ra \hat{H}$  which is  the composite map.
This may be expressed in $m \times m$ matrix form as
\be
\KK=
\pmatrix{ \KK_{11}& \cdots & \KK_{1m} \cr
                  \vdots &  \ddots & \vdots \cr
                  \KK_{m1} & \cdots  & \KK_{mm} }
 \ee
 where  
 \be
\KK_{ij} := \sum_{a=1}^N |\psi_a^{(i)} ><\phi_a^{(j)}|  .
 \ee
 We also define the linear operator $\check{\KK}:\hat{H} \ra \hat{H}$ as
 \be
 \check{\KK} := \KK - \WW 
 \ee
 and let $\II: \hat{\HH} \ra\hat{\HH}$ denote the identity map.
 Finally define, as in (\ref{Gab}) the $N \times N$ matrix  $G$ whose
 components $G_{ab}$ are the scalar products
\be
G_{ab} := <\psi^{(1)}_a| (1-\rho_1) w^*_{2,1} (1-\rho_2 )\cdots  (1-\rho_{m-1})w^*_{m+1,m}
(1-\rho_m) |\phi^{(m)}_b> .
\ee
The main result of this section is the following identity
\begin{proposition}
\be
\det (G) = \det (\II - \check{\KK}\circ \rho) .
\ee
\label{detId}
\end{proposition}

\noindent{\bf Proof:}
We begin by defining $m-1$ modified  linear maps
\be
\tilde{w}_{j+1, j}: \hat{H}_j \ra  \hat{H}_{j+1},  \quad j=1, \dots , m-1
\ee
by composition with $(1-\rho_j):\hat{H}_j \ra \hat{H}_j$
\be
\tilde{w}_{j+1, j} := w_{j+1, j} \circ (1-\rho_j)
\ee
We also define, analogously to $\NN$ and $\WW$,  the operators $\tilde{\NN}$ 
and $\tilde{\WW}$ by
\be
\tilde{\NN} := \NN \circ  (\II-\rho) =
\pmatrix{
0 & 0 & 0 & \cdots & 0&0 \cr
\tilde{w}_{21} & 0 & 0 & \cdots  & 0& 0\cr 
0 & \tilde{w}_{32}  & 0 &  \cdots & 0& 0 \cr
\vdots & \vdots & \vdots & \ddots  &\vdots &\vdots\cr
\vdots & \vdots & \vdots & \cdots  & 0 &0\cr
0 & 0 & 0 & \cdots & \tilde{w}_{m, m-1} & 0
}
\ee
\be
\tilde{\WW} :=\tilde{\NN}\circ(\II-\tilde{\NN})^{-1} =
\pmatrix{
0 & 0 & 0 & \cdots & 0 &0 \cr
\tilde{w}_{21} & 0 & 0 & \cdots  &  0& 0\cr 
\tilde{w}_{31} & \tilde{w}_{32}  & 0 &  \cdots &0& 0 \cr
\vdots & \vdots & \vdots & \ddots  &\vdots &\vdots \cr
\vdots & \vdots & \vdots & \cdots  & 0 & 0 \cr
\tilde{w}_{m1} & \tilde{w}_{m2} & \tilde{w}_{m3} & \cdots & \tilde{w}_{m, m-1} & 0 
} ,
\ee
where
\be
\tilde{w}_{k j} := \tilde{w}_{k,k-1} \circ \dots \circ \tilde{w}_{j+1,j}, \quad  m\ge k > j \ge 1.
\label{tildewij}
\ee
The dual operators are
\be
\tilde{\NN}^* = 
\pmatrix{
0 & \tilde{w}^*_{21}&0& \cdots & \cdots &0 \cr
0 & 0 & \tilde{w}^*_{32} & \cdots  &  \cdots&0 \cr 
0& 0  & 0 &  \cdots & \cdots& 0  \cr
\vdots & \vdots & \vdots & \ddots  &\vdots &\vdots \cr
0&0 & 0 & \cdots  & 0 & \tilde{w}^*_{m, m-1}\cr
0& 0 & 0 & \cdots & 0 & 0
}
\ee
and
\be
\tilde{\WW}^* = 
\pmatrix{
0 & \tilde{w}^*_{21}& \tilde{w}^*_{31} & \cdots & \cdots &\tilde{w}^*_{m1}\cr
0 & 0 & \tilde{w}^*_{32} & \cdots  &  \cdots&\tilde{w}^*_{m2}\cr 
0& 0  & 0 &  \cdots & \cdots& \tilde{w}^*_{m3} \cr
\vdots & \vdots & \vdots & \ddots  &\vdots &\vdots \cr
0&0 & 0 & \cdots  & 0 &\tilde{w}^*_{m, m-1}\cr
0& 0 & 0 & \cdots & 0 & 0
}.
\ee
Note now that
\bea
\left((\II+\tilde{\WW}^*)\circ  (\II-\rho)\right) _{1m} &\&=\pi_1(\II+ \tilde{\WW}^*)\circ  (\II-\rho)\iota_m =   \tilde{w}^*_{2,1} \circ \dots \circ 
 \tilde{w}^*_{m, m-1} \circ (1-\rho_m) \cr
 &\& =  (1-\rho_1) w^*_{2,1} (1-\rho_2 )\cdots  (1-\rho_{m-1})w^*_{m+1,m} (1-\rho_m) ,
 \eea
 and hence
 \bea
 G_{ab} &\&= <\psi_a^{(1)}| \pi_1(\II+\tilde{\WW}^*)\circ  (\II-\rho) \iota_m|\phi_b^{(m)}> \cr
 &\& = <\psi_a^{(1)}| \pi_1 (\II- (\II-\rho)\circ \NN^*)^{-1}\circ  (\II-\rho) \iota_m|\phi_b^{(m)}> .
 \eea
 From (\ref{psiaWphib}) it follows that
 \be
 G_{ab} = \delta_{ab} - \gamma_{ab} ,
 \ee
 where
 \bea
 \gamma_{ab} &\&:= -<\psi_a^{(1)}| \pi_1\left((\II- (\II-\rho)\circ\NN^*)^{-1}\circ  (\II-\rho) 
 -(\II-\NN^*)^{-1}\right)\iota_m|\phi_b^{(m)}>   \cr
&\&= <\psi_a^{(1)}| \pi_1\left((\II- (\II-\rho)\circ\NN^*)^{-1}\circ  \rho \circ (\II-\NN^*)^{-1}\right)\iota_m|\phi_b^{(m)}>
\cr
&\& = \Psi_a^T \left((\II-\NN^*)\circ (\II- (\II-\rho)\circ\NN^*)^{-1}\circ  \rho \right)\Phi_b
 \eea
 (where (\ref{PsiaT}), (\ref{Phia}) have been used in the last line). Now applying the well-known identity
 \be
 \det(\II - A\circ B) =  \det(\II - B\circ A) 
 \ee
 to the operators
 \bea
 B &\&:=  \rho\circ \Phi: \Cb^N \ra \hat{H} \cr
 A &\& := \Psi^T \circ \left((1-\NN^*)\circ (\II- (\II-\rho)\circ\NN^*)^{-1}\right) : \hat{H} \ra \Cb^N, 
 \eea
 we obtain
 \bea
 \det(G) &\&= \det \left((\II -  \rho \circ \Phi \circ \Psi^T  (\II-\NN^*)\circ (\II- (\II-\rho) \circ \NN^*)^{-1}\right) \cr
 &\& = \det \left(\II -   (\II-  \NN \circ (\II-\rho))^{-1}\circ (\II-\NN) \circ \KK \circ \rho \right) \cr
 &\& =  \det \left(  (\II-  \NN \circ (\II-\rho) - (\II-\NN) \circ \KK \circ \rho \right) \cr
  &\& =  \det \left(  (\II-  (\II-\NN) ^{-1}\circ \NN \circ \rho) -\KK \circ \rho \right)  \cr
  &\& = \det \left(\II- (\KK -\WW) \circ \rho \right) =  \det \left(\II-  \hat{\KK} \circ \rho \right),
 \eea
 where in the third and fourth lines we have used the fact that
 \be
 \det \left(\II-  \NN \circ (\II-\rho)\right) =   \det \left(\II-  \NN \right)  =1.
\ee

\section{Examples}
\subsection{Classical Lie algebra chains}
As furthert applications of the above identity, we consider coupled chains of $m$ random
matrices having values in the Lie algebras of the compact forms of the classical Lie groups
\cite{PFEDFZ}, with Itzykson-Zuber nearest neighbour exponential couplings. The partition function is of the form
\be
\Zb_N^m = \prod_{j=1}^m\left(\int dA_j \right) \prod_{j=1}^m e^{-\tr V_j(A_j)} \prod_{j=1}^{m-1}e^{\tr A_j A_{j+1}}
\label{ZNmuN}
\ee
where $\{A_j\in \grg\}_{j=1, \dots m}$  for any of the classical Lie algebras $\grg=\mathfrak{u}(N)$, $\gro(2N)$,  $\gro(2N+1)$ or $\grsp(N)$ and $dA_j $ denotes Lebesgue measure on each of these. Using the Harish-Chandra identity to integrate out the angular variables reduces these to integrals over the Cartan subalgebras. For the case $\gru(N)$, identified as Hermitian matrix chains,  this gives the usual Itzykson-Zuber reduced integral \cite{IZ, EM}
\be
\Zb_N^m \propto \prod_{j=1}^m \left( \int_{\Rb^N}\prod_{a=1}^N dx_a^{(j)}  e^{-V_j(x_a^{(j)})}\right)
\Delta(x_1^{(1)}, \dots, x_N^{(1)}) \Delta(x_1^{(m)}, \dots, x_N^{(m)})
 \prod_{j=1}^{m-1} \det (e^{x_a^{(j)}x_b^{(j)}})
 \label{ZNmuNred}
\ee
where
\be
\Delta(x_1, \dots, x_N) := \det(x_b^{N-a})|_{1\le a,b \le N} =\prod_{1\le a < b \le N}(x_a-x_b)
\ee
is the Vandermonde determinant. 
For the other classical Lie algebras it gives (see \cite{PFEDFZ} for the case $m=2$)
\smallskip

\noindent $\gro(2N)$:
\bea
\Zb_N^m &\&\propto \prod_{j=1}^m \left( \int_{\Rb^N}\prod_{a=1}^N dx_a^{(j)}  e^{-V_j(x_a^{(j)})}\right)
\Delta((x_1^{(1)})^2, \dots, (x_N^{(1)})^2) \Delta((x_1^{(m)})^2, \dots, (x_N^{(m)})^2) \cr
 &\& {\hskip 150 pt} \times  \prod_{j=1}^{m-1}    \det (2\cosh(x_a^{(j)}x_b^{(j)}))
  \label{ZNmo2N}
\eea

\noindent $\gro(2N+1)$ or $\grsp(N)$:
\bea
\Zb_N^m \propto &\&\prod_{j=1}^m \left( \int_{\Rb^N}\prod_{a=1}^N dx_a^{(j)}  e^{-V_j(x_a^{(j)})}\right)
\left(\prod_{a=1}^Nx_a^{(1)}\right)\Delta((x_1^{(1)})^2, \dots, (x_N^{(1)})^2) \cr
  &\& \times  \left(\prod_{a=1}^Nx_a^{(m)}\right) \Delta((x_1^{(m)})^2, \dots, (x_N^{(m)})^2) 
 \prod_{j=1}^{m-1}    \det (2\sinh(x_a^{(j)}x_b^{(j)})),
   \label{ZNmo2N1}
\eea
where the potentials $\{V_j\}_{j=1, \dots, m}$ must be chosen as even functions of their
arguments. 

Replacing the monomial elements $\left((x^{(1)}_b)^{N-a}, \ (x^{(m)}_b)^{N-a}\right)$  in the Vandermonde determinants  in (\ref{ZNmuNred}) by $\left(\psi_a(x_b^{(1)}),\  \phi_a(x_b^{(m)})\right)$, where
 \bea
 \psi_a(x^{(1)})&\&:= p_{a-1}(x^{(1)})e^{- {1\over 2}V_1(x^{(1)})}, \cr
  \phi_a(x^{(m)})&\& := s_{a-1}(x^{(m)})e^{- {1\over 2}V_m(x^{(m)})}, \quad a=1, \dots N
  \label{psiphia}
  \eea
and $p_{a}(x^{(1)}),\ s_{a}(x^{(m)})$ are polynomials of degree $a\in \Nb$ 
  satisfying the biorthogonality conditions
\be
 \prod_{j=1}^m\left(\int_{\Gamma_j} dx^{(j)} e^{-V_j(x^{(j)})}\right)
p_a(x^{(1)})s_b(x^{(m)}) \prod_{j=1}^{m-1}e^{x^{(j+1)}x^{(j)}}
=\delta_{ab},
\ee
we obtain the joint probability density (\ref{PNm}) with 
$(\Gamma_j, d\mu_j) = (\Rb, dx^{(j)}), j=1, \dots, m$ and
\be
w_{j+1, j}(x^{(j+1)}, x^{(j)}) := e^{x^{(j)} x^{(j+1)} }.
\ee
Thus, applying Proposition \ref{detId} to this case, and choosing the functions
$\rho_j$ to be the characteristic functions $\chi_{J_j}$ of unions of subintervals $J_j =\bigcup_l J_{jl}\subset \Rb$  gives the gap probability
\be
    P^{N,m}_{\chi_{J_1}, \dots, \chi_{J_m}}: =  E^{N,m}(0, {\bf J}) =C_{Nm} \det (\II - \check{\KK} \circ \chi_{\bf J})
\ee
  where the $m\times m$ matrix Fredholm operator $\check{\KK}$ is defined as in  (\ref{Kcij}), (\ref{Kij}) 
  with  $(\psi_a^{(i)},   \phi_a^{(i)})$ defined in (\ref{wij}), (\ref{psiaj}), (\ref{phiaj}).
  
  The other two cases are nearly identical, the only difference being that, for $\gro(2N)$,
  the potentials $V_j(x^{(j)})$ should be taken as even functions, and
  the  biorthogonal polynomials $(p_{a-1}(x^{(1)}) s_{a-1}(x^{(m)}) )$ in (\ref{psiphia})
  replaced by even polynomials of even degrees
   \bea
 \psi_a(x^{(1)})&\&:=p_{2(a-1)}(x^{(1)})e^{- {1\over 2}V_1(x^{(1)})},  \cr
  \phi_a(x^{(m)})&\& :=s_{2(a-1)}(x^{(m)})e^{- {1\over 2}V_m(x^{(1)})}, \quad a=1, \dots N
  \label{psiphiaoN}
  \eea
  satisfying the biorthogonality conditions
\be
 \prod_{j=1}^m\left(\int_{\Gamma_j^N} dx^{(j)} e^{-V_j(x^{(j)})}\right)
p_{2a}(x^{(1)})s_{2b}(x^{(m)}) \prod_{j=1}^{m-1}2\cosh(x^{(j+1)}x^{(j)})
=\delta_{ab},
\ee
The functions $w_{j+1, j}$ entering in   (\ref{wij}), (\ref{psiaj}), (\ref{phiaj}) are
\be
w_{j+1, j} (x^{(j+1)},  x^{(j)}) := 2\cosh(x^{(j+1)}x^{(j)}).
\ee

For $\gro(2N+1)$ and $ \grsp(N)$  the potentials $V_j(x^{(j)})$ must again be taken as even functions,
  the  biorthogonal polynomials $(p_{a-1}(x^{(1)}) s_{a-1}(x^{(m)}) )$ in (\ref{psiphia})
  are replaced by odd polynomials of odd degrees
   \bea
 \psi_a(x^{(1)})&\&:=p_{2a-1}(x^{(1)})e^{- {1\over 2}V_1(x^{(1)})},  \cr
  \phi_a(x^{(m)})&\& :=s_{2a-1}(x^{(m)})e^{- {1\over 2}V_m(x^{(1)})}, \quad a=1, \dots N
  \label{psiphiaoN}
  \eea
  satisfying the biorthogonality conditions
\be
 \prod_{j=1}^m\left(\int_{\Gamma_j^N} dx^{(j)} e^{-V_j(x^{(j)})}\right)
p_{2a-1}(x^{(1)})s_{2b-1}(x^{(m)}) \prod_{j=1}^{m-1}2\sinh(x^{(j+1)}x^{(j)})
=\delta_{ab},
\ee
and the functions $w_{j+1, j}$ entering in   (\ref{wij}), (\ref{psiaj}), (\ref{phiaj}) are
\be
w_{j+1, j} (x^{(j+1)},  x^{(j)}) := 2\sinh(x^{(j+1)}x^{(j)}).
\ee

\br In view of the fact that the joint probability densities for  the cases  $\gro(2N)$,  $\gro(2N+1)$ and  $\grsp(N)$ are all even functions of their arguments, if  averages only over even functions are considered, as would be the case, for example, in computing expectation values of conjugation invariant functions of the matrices, the coupling terms $2\cosh(x^{(j)} x^{(j+1)})$ and $2\sinh(x^{(j)} x^{(j+1)})$ appearing may equivalently
by replaced by the usual exponential couplings $e^{x^{j} x^{j+1}}$.
\er
\subsection{Nonexponential couplings}

As explained in  \cite{HO},  Appendix A, another generalization of the Hermitian
 matrix chain can be made  by choosing a set of functions $f_j(z)$ having a Taylor series 
 expansion about the origin of the form
 \be
f_j(x)= 1+ \sum_{k=1}^\infty r_j(1) \cdots r_j(k) x^k , \quad j=1, \dots, m-1.
\ee
The exponential coupling term $e^{\tr A_j A_{j+1}}$  in (\ref{ZNmuN})  
may be replaced by  the more general set of conjugation invariant functions 
\be
\tau_j( A_j A_{j+1}) := \sum_{\lambda \atop  \ell(\lambda)\le N}
\left( \prod_{i,k \in \lambda}r_j(N+k-i)\right) d_{\lambda,N}  s_\lambda(A_jA_{j+1}),
 \label{tau_j}
 \ee
where $s_\lambda$ is the Schur function corresponding to  partition $\lambda= (\lambda_1\ge \cdots \ge \lambda_{\ell(\lambda)} >0)$  of length $\ell(\lambda)$,  viewed as a character, 
and  $d_{\lambda,N}$ is the dimension of the irreducible tensor representation 
of $\grGl(N)$ corresponding to $\lambda$.
  The joint probability density for the eigenvalues on the chain is then again of the form  (\ref{PNm})
  with the  functions $(\psi_a(x^{(1)}), \phi_a(x^{(m)}))$ defined as in (\ref{psi1phimpolynom}) satisfying the biorthogonailty relations  (\ref{biorthog}), but with the functions $w_{j+1,j}$ defined as
  \be
  w_{j+1,j}(x^{(j+1)}, x^{(j)}) := f_j (x^{(j)} x^{(j+1)}).
  \ee
  
  An example of such a coupling is obtained by choosing a set of $m-1$ pairs of constants
  $\{(a_1,z_1), \dots ,   (a_{m-1},z_{m-1} )\}$ and setting
  \be
  f_j(x):=  (1 - z_j x)^{N-a_j-1},  \quad j=1, \dots, m-1.
  \ee
  Hence
  \be 
  w_{j+1, j} (x^{(j+1)}, x^{(j)}) =   (1 - z_j x^{(j)}x^{(j+1)})^{N-a_j-1},
  \ee
which corresponds to
  \be
  r_j(i) = {z_j (a_j -N +i) \over i}.
  \ee
   This choice gives
  \be
  \tau_j(A_j A_{j+1}) = \det(\II - z_jA_j A_{j+1})^{-a_j}
  \ee
  and hence the partition function (\ref{ZNmuN}) for the coupled random matrix chain is 
  \be
\Zb_N^m = \prod_{j=1}^m\left(\int dA_j \right) \prod_{j=1}^m e^{-\tr V_j(A_j)} \prod_{j=1}^{m-1}\det(\II - z_jA_j A_{j+1})^{-a_j}.
\label{ZNmuNdet}
\ee

All the above formulae for multi-point correlators, gap probabilities and Janossy 
densities are applicable to these cases, simply by adapting the general fomulae
to these choices for the coupling functions  $w_{j+1, j} $.

\bigskip\bigskip \noindent{\it Acknowledgements.}
The authors  would like to thank A. Prats-Ferrer for helpful comments regarding the
case of coupled classical Lie algebra chains.
 \bigskip \bigskip



\begin{thebibliography}{99}

\bibitem{A} C.  Andr\'eief, ``Note sur une relation pour les int\'egrales d\'efinies des
produits des fonctions'', {\em M\'em. Soc. Sci. Bordeaux}, {\bf 2} 1--14 (1883).

\bibitem{EM} B. Eynard  and M.L. Mehta, ``Matrices coupled in a chain:
eigenvalue correlations'', {\em J. Phys. A} {\bf 31}, 4449--4456 (1998).

\bibitem{H} J. Harnad, ``Janossy densities, multimatrix spacing distributions and Fredholm resolvents'',
 {\em Int. Math. Res. Not. } {\bf 48 }, 2599--2609 (2004).
 
 \bibitem{HO} J. Harnad and Alexander Yu. Orlov, ``Fermionic construction of partition functions for two-matrix models and perturbative Schur function expansions '',
 {\em J. Phys.  A} {\bf 39}, 8783-8809 (2006) .
 
 \bibitem{IZ} C. Itzykson and J.B. Zuber, ``The planar approximation  (II)'', {\em J. Math. Phys.} 
 {\bf 21}, 411--421 (1980).
 
 \bibitem{PS}  M. Pr\"ahofer and H. Spohn, ``Scale invariance of the PNG droplet and 
the Airy process'', {\em J. Stat. Phys.} {\bf 108,} 1071--1106 (2002).

 \bibitem{PFEDFZ}  A. Prats-Ferrer, B. Eynard, P. di Francesco and  J.-B. Zuber, ``Correlation Functions of Harish-Chandra Integrals over the Orthogonal and Symplectic Groups'', {\em J. Stat. Phys.} {\bf 129}, 885-935 (2007).

\bibitem{S} Alexander Soshnikov,  ``Janossy densities of coupled random matrices'',
{\em Commun. Math. Phys.}  {\bf 251}, 447-471  (2004) .
 
 \bibitem{TW1} Craig A. Tracy  and Harold Widom, `Correlation Functions, Cluster Functions and Spacing Distributions for Random Matrices'',  {\em J. Stat. Phys.} {\bf 92}, 809-835 (1998).
 
\bibitem{TW2} Craig A. Tracy  and Harold Widom, ``Differential equations for Dyson processes'',
 {\em Commun. Math. Phys.} {\bf 252 }, 7--41 (2004) .




\end{thebibliography}
\end{document}